\documentclass{article}

\usepackage{microtype}
\usepackage{graphicx}
\usepackage{subfigure}
\usepackage{booktabs} % for professional tables

\usepackage{forest}
\useforestlibrary{edges}
\usepackage{tikz}

\usepackage{array}

\usepackage{hyperref}

\usepackage[accepted]{icml2025}

\usepackage{amsmath}
\usepackage{amssymb}
\usepackage{mathtools}
\usepackage{amsthm}

\usepackage[capitalize,noabbrev]{cleveref}

\theoremstyle{plain}

\theoremstyle{definition}

\theoremstyle{remark}

\usepackage[textsize=tiny]{todonotes}

\icmltitlerunning{Wearable AI in the Era of Large Sensor Models}

\begin{document}

\twocolumn[

\icmltitle{Wearable AI in the Era of Large Sensor Models}

% It is OKAY to include author information, even for blind
% submissions: the style file will automatically remove it for you
% unless you've provided the [accepted] option to the icml2025
% package.

% List of affiliations: The first argument should be a (short)
% identifier you will use later to specify author affiliations
% Academic affiliations should list Department, University, City, Region, Country
% Industry affiliations should list Company, City, Region, Country

% You can specify symbols, otherwise they are numbered in order.
% Ideally, you should not use this facility. Affiliations will be numbered
% in order of appearance and this is the preferred way.
\icmlsetsymbol{equal}{*}

\begin{icmlauthorlist}
\icmlauthor{Yize Cai}{1}
\icmlauthor{Baoshen Guo}{2}
\icmlauthor{Guobin Shen}{1}
\icmlauthor{Zhiqing Hong}{1}
\end{icmlauthorlist}

% \icmlaffiliation{yyy}{Department of XXX, University of YYY, Location, Country}
% \icmlaffiliation{comp}{Company Name, Location, Country}
% \icmlaffiliation{sch}{School of ZZZ, Institute of WWW, Location, Country}

% \icmlcorrespondingauthor{Firstname1 Lastname1}{first1.last1@xxx.edu}
% \icmlcorrespondingauthor{Firstname2 Lastname2}{first2.last2@www.uk}

\icmlaffiliation{1}{The Hong Kong University of Science and Technology (Guangzhou)}
\icmlaffiliation{2}{Singapore-MIT Alliance for Research and Technology}
\icmlcorrespondingauthor{Zhiqing Hong}{zhiqinghong@hkust-gz.edu.cn}

% You may provide any keywords that you
% find helpful for describing your paper; these are used to populate
% the "keywords" metadata in the PDF but will not be shown in the document
\icmlkeywords{Machine Learning, ICML}

\vskip 0.3in
]

% this must go after the closing bracket ] following \twocolumn[ ...

% This command actually creates the footnote in the first column
% listing the affiliations and the copyright notice.
% The command takes one argument, which is text to display at the start of the footnote.
% The \icmlEqualContribution command is standard text for equal contribution.
% Remove it (just {}) if you do not need this facility.

\printAffiliationsAndNotice{}  % leave blank if no need to mention equal contribution
% \printAffiliationsAndNotice{\icmlEqualContribution} % otherwise use the standard text.

\begin{abstract}
% such as inertial measurement units (IMU), electrocardiography (ECG), and Global Navigation Satellite Systems (GNSS)

% In recent years, human-centric wearable artificial intelligence (AI) has advanced rapidly, leveraging wearable sensing modalities to understand human physiology and behavior. 
As an effective approach to understanding the human-centric physical world, Wearable Artificial Intelligence (AI), which leverages multimodal wearable sensors to understand human physiology and behavior, has attracted increasing attention in recent years.
However, existing sensor models remain largely siloed by modality and task, lacking a unified paradigm for integrating diverse wearable modalities, training strategies, and achieving robust generalization in real-world applications.
Motivated by the success of multimodal foundation models, which learn transferable representations from massive multimodal data, we argue that Large Sensor Models (LSMs), defined as foundation models trained on large-scale and multimodal wearable data, offer a promising pathway toward a more general and scalable framework for wearable AI.
In this position paper, we formalize the data substrate underlying LSMs, analyze the unique challenges of large-scale wearable sensing, and articulate two directions: (i) LSMs without language capability and (ii) LSMs with language capability. % each suited to different deployment and interaction scenarios. 
We further discuss representative application areas that can be unlocked by such models. 
Through this paper, we encourage the community to explore LSMs as a foundational approach for the next generation of human-centric AI systems. 
\end{abstract}

\section{Introduction}

\begin{figure}[t]
	\centering
	\includegraphics[width=3.3in]{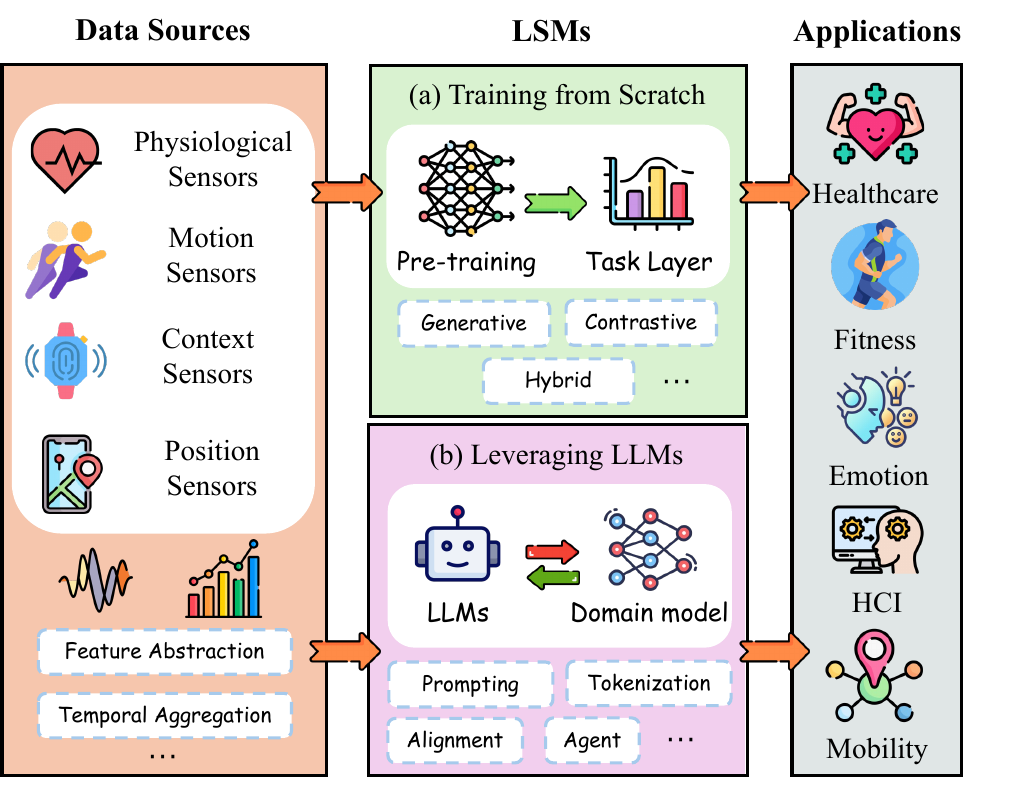}
	\caption{Overview of LSMs. (a) LSMs without language capability. (b) LSMs with language capability.}
	\label{fig:lsm}
\end{figure}

Wearable Artificial Intelligence (AI) systems aim to understand and respond to individual contexts and preferences, allowing technology to function as a natural extension of the human self~\cite{abd2023systematic}.
Physiological, physical, and behavioral signals provide rich and fine-grained evidence about the state of the body, offering models access to patterns of movement, cardiac dynamics, and environmental interaction.
During the past decade, a substantial body of research has focused on leveraging one or multiple sensing modalities~\cite{he2023toward, gu2025foundation}, such as inertial measurement units (IMU), electrocardiography (ECG), and Global Navigation Satellite Systems (GNSS), to design end-to-end models tailored to specific downstream tasks, such as human activity recognition (HAR) and emotion recognition (ER). 
While these approaches have achieved strong task-level performance, they remain largely specialized and fragmented across modalities and applications.
 
Wearable data exhibit substantial cross-modality heterogeneity, intra-modality distribution shifts, high noise levels, and irregular sampling patterns, making it fundamentally challenging for small-scale models trained on limited data to generalize effectively~\cite{cai2025towards}.
Meanwhile, the widespread adoption of consumer and industrial wearable devices has led to an unprecedented growth of multimodal sensor data. 
Furthermore, emerging wearable-driven applications are rapidly expanding, increasing the diversity of required tasks~\cite{haresamudram2025past}. 
The prevailing strategy of training separate models for each modality and task results in high development and deployment costs, limited knowledge transfer, and poor scalability. 
In recent years, foundation models (FMs)~\cite{bommasani2021opportunities}, trained on large-scale data with advanced architectures and designed to support various downstream tasks, have reshaped research paradigms in language, vision, and multimodal learning~\cite{zhang2024mm}. 
Their core principle of learning transferable representations through scale aligns naturally with the growing task demands and data availability in Wearable AI, suggesting a promising direction for building unified and scalable sensor models.

In this position paper, we propose \textbf{Large Sensor Models (LSMs)} as universal multimodal FMs tailored for the wearable sensing domain. 
LSMs ingest heterogeneous sensor data and learn unified representational substrates that remain temporally faithful to underlying human behaviors and environmental context. 
Such representations are expected to align across modalities, remain robust under distribution shifts and dynamic modality combinations, and exhibit predictable performance gains as data volume and model capacity expand. 
The goal of LSMs is to provide transferable semantic and unified representations that can be readily adapted across human-centric tasks and various sensor data distributions. 
By mapping continuous and noisy sensor streams into composable higher-level abstractions, LSMs aim to consolidate fragmented wearable AI efforts into a coherent and scalable research agenda, thereby positioning themselves as the sensor-domain counterpart of FMs in language and vision and charting a path toward the systematic evolution of pervasive sensor intelligence.

\textbf{Research Agenda.} In this position paper, we present key insights that lay the foundation for a principled agenda on LSMs, as shown in Figure~\ref{fig:lsm}.

\noindent \textbf{(1) Data Substrate} (Section~\ref{sec:data}). 
Wearable sensing spans various physiological sensors, motion sensors, context sensors, and position sensors, each characterized by heterogeneous temporal dynamics, variable sampling rates, and distinct noise profiles. 
The central challenge is to learn representations that remain faithful to underlying physical properties while being robust to distribution shifts and measurement imperfections caused by diverse modalities, users, device types, and sensor placements.

\noindent \textbf{(2) Modeling Paradigms} (Section~\ref{sec:modeling1} and~\ref{sec:modeling2}). 
We present two development directions for LSMs:
\textit{(i) LSMs without language capability} focus on scaling and systematizing pretrained sensor models through increased model capacity, expanded multimodal data, and advanced modeling strategies. 
\textit{(ii) LSMs with language capability} integrate large language models (LLMs) into the sensor modeling pipeline, enabling structured reasoning, semantic abstraction, and cross-modal alignment.
This direction leverages the emergent reasoning abilities of LLMs to enhance interpretability, interaction, and decision support in sensor driven systems.

\noindent \textbf{(3) Application Areas} (Section~\ref{sec:application}).
By unifying heterogeneous sensor streams into embeddings, LSMs open the path toward universal sensor intelligence across domains, such as healthcare monitoring, fitness and lifestyle, embodied intelligence, emotion support, and transportation and mobility.  
Such models have the potential to support adaptive, context-aware, and privacy conscious applications that operate efficiently under real-world computational constraints.

By foregrounding these three dimensions, we argue for a cohesive research paradigm that situates LSMs as the next step in the evolution of wearable AI.

\textbf{Contributions:} This paper makes the following contributions: \textbf{(1) A paradigm shift for wearable AI.} 
We position LSMs as principled FMs for multimodal sensor data, reframing wearable AI not as a collection of task-specific pipelines, but as a scalable, unified paradigm. 
\textbf{(2) A structured research agenda.}
We critically synthesize preliminary efforts toward LSMs, systematize progress across data organization, modeling paradigms, and application domains. 
In doing so, we provide a structured roadmap that organizes scattered progresses into a coherent trajectory.
\textbf{(3) A clarification of future directions.} 
We systematically examine the challenges of data and modeling paradigms, and outline potential application opportunities, highlighting promising directions for future LSM development.

\section{Background}

\begin{figure*}[t]
	\centering
	\includegraphics[width=6in]{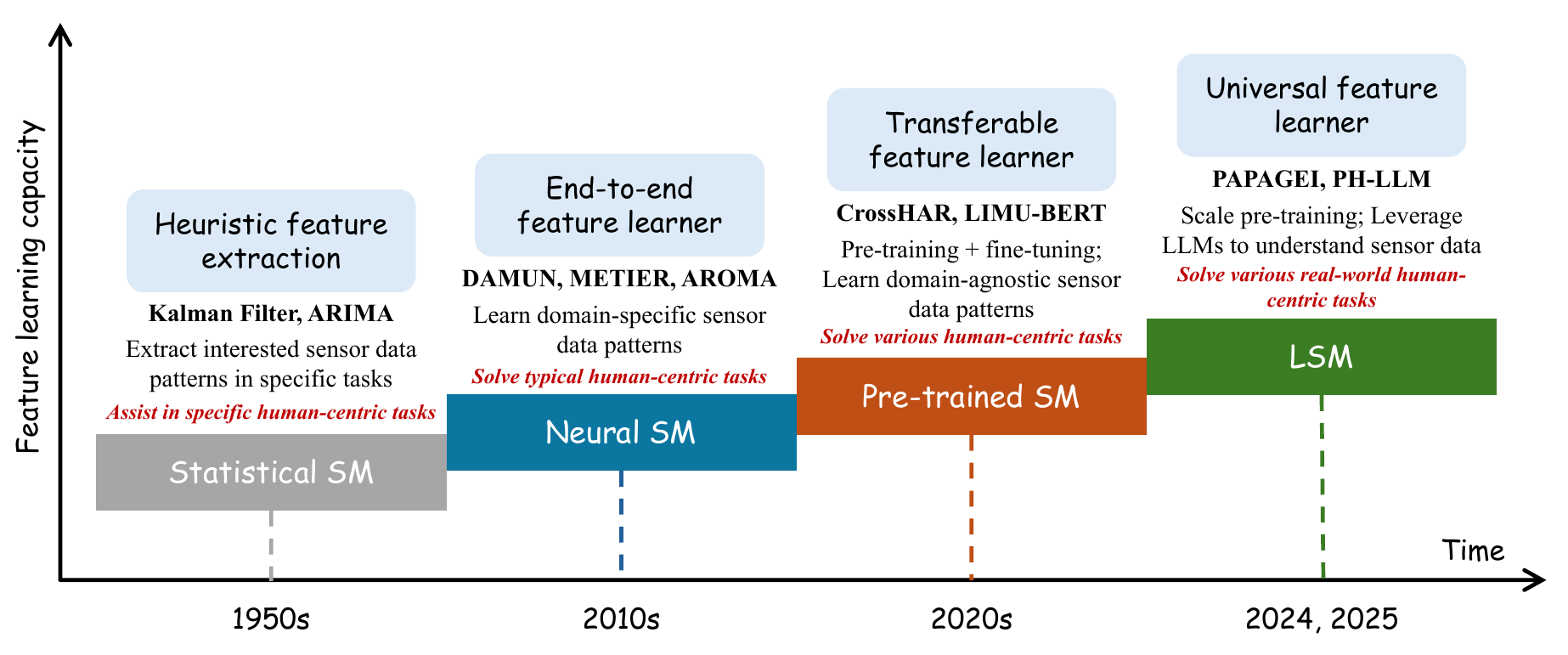}
	\caption{A roadmap of the four generations of sensor models (SMs) from the perspective of feature learning capacity.}
	\label{fig:roadmap}
\end{figure*}

\subsection{Foundation Models}

Foundation models (FMs) have transformed the landscape of machine learning by demonstrating that large-scale pretraining can yield general-purpose representations with strong transferability and generalization capability~\cite{bommasani2021opportunities}, such as LLaMA~\cite{touvron2023llama} and GPT-4~\cite{achiam2023gpt} in natural language processing (NLP), ViT~\cite{dosovitskiy2020image} and DINOv2~\cite{oquab2023dinov2} in computer vision (CV), and CLIP ~\cite{radford2021learning} and multimodal large language models (MLLMs)~\cite{zhang2024mm, yin2024survey} in multimodal models.
More recently, advanced time-series methods such as TimesFM~\cite{das2024decoder}, Chronos~\cite{ansari2024chronos}, and TimeGPT~\cite{garza2023timegpt} have extended FMs to time-series analysis, demonstrating that general time-series data can also support general-purpose modeling under sufficient scale.

Taken together, these developments reveal a common methodological principle: scaling data and model capacity yields emergent generalization, robust transfer, and adaptability across diverse downstream tasks. 
However, wearable sensing introduces distinctive challenges that fundamentally differentiate LSMs from prior FMs. 
Unlike the above modalities, wearable data consist of continuous, noisy, and heterogeneous multimodal streams characterized by irregular sampling, distribution shifts, and embodied physical constraints.
These conditions demand models that are modality-robust and temporally faithful, generalizable under distribution shifts, and deployable under privacy-preserving and resource-constrained settings. 
LSMs, therefore, inherit the scaling philosophy of FMs, while requiring new architectural and training principles tailored to the pervasive and embodied nature of human-centric sensor data.

\subsection{Research Roadmap of Sensor Models}

The evolution of Sensor Models (SMs) can be organized into four stages: \textit{(1) statistical SMs}, \textit{(2) neural SMs}, \textit{(3) pre-trained SMs}, and \textit{(4) LSMs}, as shown in Figure~\ref{fig:roadmap}. 
Statistical SMs are typically constructed based on manually engineered features and predefined inductive assumptions, such as Kalman Filter~\cite{mazza2012optimized} and ARIMA~\cite{kong2022time}.
Neural SMs advance representation learning from raw sensor data and improve accuracy, such as DAMUN~\cite{bai2020adversarial}, METIER~\cite{chen2020metier}, and AROMA~\cite{peng2018aroma}. 
Pre-trained SMs mitigate label scarcity by learning representations from unlabeled data, improving label efficiency and robustness, such as CrossHAR~\cite{hong2024crosshar}, WEPOS~\cite{guo2022wepos}, and LIMU-BERT~\cite{xu2021limu}.
LSMs follow naturally and extend the established trajectory in language, vision, and multimodality models.
We posit that LSMs provide development resolutions through two strategies: (i) systematically exploring the scaling effect on model performance, such as ECG-FM~\cite{mckeen2024ecg}, and
(ii) leveraging the ability of LLMs to intuitively understand the physical world to advance LSMs, such as PH-LLM~\cite{khasentino2025personal}.
These strategies position LSMs as principled FMs that unify sensor data understanding and enable robust performance in human-centered sensing.

\section{Wearable Data Substrate}
\label{sec:data}

\begin{table*}[t]
\centering
\small
\caption{A summary of representative sensor datasets.}
\scalebox{0.8}{
\begin{tabular}{@{} m{2.5cm}<{\centering} | m{0.8cm}<{\centering} m{0.8cm}<{\centering} m{0.8cm}<{\centering} m{0.8cm}<{\centering} m{2.2cm}<{\centering} | m{3.4cm}<{\centering} m{4cm}<{\centering} m{1.8cm}<{\centering} @{}}
\toprule
Dataset    & Phys.      & Mot.                      & Cont.                      & Pos.                      & Source   & Modality & Label / Information & \# Subjects \\ \midrule
\href{https://figshare.scilifelab.se/articles/dataset/CODE_dataset/15169716}{CODE-15} & \checkmark &                           &                           &                           & clinical & ECG      & disease and mortality labels & 234K \\
\href{https://zenodo.org/records/4905618}{SaMi-Trop} & \checkmark &                           &                           &                           & clinical & ECG      & mortality labels & 2K \\
\href{https://zenodo.org/records/8393007}{IKEM} & \checkmark &                           &                           &                           & clinical & ECG      & -     & 30K \\
\href{https://physionet.org/about/database/}{PhysioNet} & \checkmark &                           &                           &                           & clinical & ECG      & disease label and report & 45K \\
\href{https://sleepdata.org/}{NSRR} & \checkmark &                           &                           &                           & clinical & PSG      & sleep reports  & 27K \\
\href{https://vitalvideos.org/}{VV} & \checkmark &                           &                           &                           & clinical & PPG      & blood pressure labels & 850 \\
\href{https://figshare.com/articles/dataset/PPG-BP_Database_zip/5459299}{PPG-BP} & \checkmark &                           &                           &                           & clinical & PPG      & cardiovascular disease labels & 219 \\
\href{https://github.com/pulselabteam/PulseDB}{PulseDB} & \checkmark &                           &                           &                           & clinical & ECG, PPG & systolic blood pressure labels & 5K \\
% SDB        & \checkmark &                           &                           &                           & clinical & PSG, PPG & sleep disordered disease labels & 160 \\
\href{https://www.sleepdata.org/datasets/mesa}{MESA} & \checkmark &                           &                           &                           & clinical & ECG, PPG, EEG & sleep labels & 2K \\
\href{https://physionet.org/content/vtac/1.0/}{VTaC} & \checkmark &                           &                           &                           & clinical & ECG, PPG, ABP & ICU alarm labels & 2K \\
\href{https://tison.ucsf.edu/ppg-diabetes}{UCSF-PPG} & \checkmark &                           &                           &                           & wearable & PPG & diabetes labels & 54K \\
\href{https://isip.piconepress.com/projects/tuh_eeg/}{TUH-EEG} & \checkmark &                           &                           &                           & clinical & EEG & seizure labels and reports & 16K \\
\href{https://github.com/facebookresearch/emg2pose}{emg2pose}   & \checkmark &                           &                           &                           & wearable & EMG & gesture labels & 193 \\
\href{https://github.com/facebookresearch/emg2qwerty}{emg2qwerty} & \checkmark &                           &                           &                           & wearable & EMG & typing labels & 108 \\
\href{https://alchemy18.github.io/FEEL_Benchmark/}{FEEL}       & \checkmark &                           &                           &                           & wearable & PPG, EDA & emotion labels & - \\
\href{https://ora.ox.ac.uk/objects/uuid:99d7c092-d865-4a19-b096-cc16440cd001}{CAPTURE-24} &                           & \checkmark &                           &                           & wearable & IMU      & activity labels & 151 \\
\href{https://www.ntnu.edu/hunt/hunt4}{HUNT4} &                           & \checkmark &                           &                           & wearable & IMU      & - & 35K \\
\href{https://www.covid-19-sounds.org/en/blog/neurips_dataset.html}{COVID-19 Sounds} &                           &       &\checkmark                 &                           & wearable & acoustic & disease labels & 36K \\
\href{https://github.com/ziruisongbest/geocomp}{GeoComp} &                           &       &                           & \checkmark                & wearable & location & geographic labels & 36K \\
% OSV-5M          &                           &       &                           & \checkmark                & wearable & location & geographic labels & 36K \\
\midrule
\href{https://github.com/Google-Health/consumer-health-research}{PH-LLM} & \checkmark & \checkmark &                           &                           & wearable & - & Q\&A for sleep and fitness & - \\
\href{https://www.ukbiobank.ac.uk/about-our-data/}{UK-Biobank} & \checkmark & \checkmark &                           &                           & both     & ECG, PPG, IMU & - & - \\
\href{https://wwwn.cdc.gov/nchs/nhanes/}{NHANES} & \checkmark & \checkmark &                           &                           & both     & - & - & - \\
\href{https://physionet.org/content/mimiciv/3.1/}{MIMIC} & \checkmark &                           & \checkmark &                           & clinical & ECG, PPG, ABP, acoustic & - & 30K \\
\href{https://vitaldb.net/dataset/}{VitalDB} & \checkmark &                           & \checkmark &                           & clinical & ECG, PPG, ABP, acoustic & surgical report & - \\
\href{https://physionet.org/content/mc-med/1.0.1/}{MC-MED} & \checkmark &                           & \checkmark &                           & clinical & ECG, PPG, ABP, acoustic & medical records & - \\
\href{https://ego4d-data.org/}{Ego4D} &            & \checkmark                & \checkmark &                           & wearable & IMU, acoustic & activity labels and captions & 923 \\
\href{https://zenodo.org/records/10407279}{EmoWear} & \checkmark & \checkmark                &            &                           & wearable & SCG(IMU), ECG, BVP, RSP, EDA, SKT & emotion labels & 49 \\
\href{http://extrasensory.ucsd.edu/}{ExtraSensory} & \checkmark &                         & \checkmark & \checkmark                & wearable & IMU, location, acoustic & activity labels & 60 \\
\href{https://zenodo.org/records/7606611}{K-EmoPhone} &              & \checkmark              & \checkmark & \checkmark                & wearable & IMU, location & smartphone usage logs, activity and emotion records & 77 \\
\href{https://osf.io/rgkvq}{TEPT} &              & \checkmark              & \checkmark & \checkmark                & wearable & ECG, location & transport contextual and emotion records & 44 \\
\bottomrule
\end{tabular}
}
\label{tab:dataset}
\end{table*}

The foundation of LSMs is data, which serves both as the substrate for representation learning and as the determinant of a model’s ability to generalize across heterogeneous sensing tasks. 
Unlike text or images, where tokens and pixels provide relatively standardized units, wearable data span multiple abstraction structures, each reflecting distinct information. 
This section introduces the main sensor modalities, articulates the unique challenges posed by wearable data, and discusses how feature abstraction and temporal granularity shape wearable data. 
A comprehensive summary of advanced wearable sensing datasets is shown in Table~\ref{tab:dataset}.

\subsection{Data Modality}

Wearable sensing spans a spectrum of modalities that reflect complementary aspects of human physiology, behavior, and context. 
We highlight four representative sensor types that together form the substrate of LSMs.

\textbf{Physiological Sensors.}
Signals such as ECG, photoplethysmography (PPG), and electrodermal activity (EDA) provide direct access to biological processes, such as cardiovascular dynamics, autonomic regulation, and stress responses~\cite{castaneda2018review}. 
These signals are low-dimensional but highly structured, governed by physiological laws that impose rhythmicity and coupling across modalities~\cite{heikenfeld2018wearable}. 
Effective modeling requires representations that preserve structural dynamics while remaining robust to motion artifacts.

\textbf{Motion Sensors.}
Accelerometers, gyroscopes, and magnetometers capture movement patterns across multiple temporal scales, from fine-grained tremors to long-duration locomotion~\cite{wang2023wearable}. 
Motion data exhibit rich variability across individuals and devices, with differences in sensor placement, calibration, and sampling rate introducing distribution shifts~\cite{cai2025towards}. 
Effective modeling requires LSMs have the ability to disentangle the invariants of human motion from domain-specific features to enable strong generalization capability.

\textbf{Context Sensors.}
Context modalities such as Bluetooth, acoustic, and ultra-wideband provide relational information that links individuals to spatial context and often serve as auxiliary information for motion modalities~\cite{ultraposer, devrio2023smartposer, demirel2025using}.
However, they are often sparse, asynchronous, and susceptible to interference, demanding representational strategies that reconcile irregular sampling with robust spatial reasoning.

\textbf{Position Sensors.}
Global navigation systems such as GPS~\cite{leick2015gps} and Beidou~\cite{yang2019introduction} track mobility in outdoor environments, offering trajectories that reflect daily routines, activity spaces, and exposure patterns~\cite{zhang2025urban, chang2021mobility}. 
Position signals operate at longer temporal horizons, yet they remain noisy and incomplete due to signal occlusion and energy-saving duty cycles.
Constructing unified representations, therefore, requires aligning coarse-grained geospatial trajectories with fine-grained sensor streams to capture compressive behavior information.

\subsection{Data Structure} 

Wearable data can be organized along two orthogonal dimensions: feature abstraction and temporal granularity.
These two dimensions jointly determine the sensor data formulation of LSMs.

\textbf{Feature Abstraction.}
The wearable data for LSMs can be organized into two levels of features.
Low-level data consist of raw sensor data that capture fine-grained physiological and behavioral dynamics. 
Most existing approaches operate at this level to detect motion events~\cite{narayanswamy2024scaling, thapa2024sleepfm} or health-related states~\cite{abbaspourazad2024wearable, abbaspourazad2024largescale}. 
Although these signals provide rich temporal detail, they are often corrupted by noise and motion artifacts.
High-level signals are derived constructs computed from raw data using validated algorithms or expert knowledge.
For example, \citet{erturk2025beyond} employ 27 expert-selected measures,~\citet{yun2024unsupervised} leverage expert-defined measures extracted from clinical records as phenotypes, and \citet{pillai2025papagei} convert PPG into predefined complementary high-level data.
Such representations are more interpretable, directly aligned with behavioral or clinical concepts, yet inevitably compress and simplify the underlying dynamics.

\textbf{Temporal Granularity.}
Temporal granularity transforms irregular and modality-specific streams into standardized summaries at predefined standards, trading temporal resolution for stability and alignment. 
Minute-level granularity preserves transient dynamics critical for event-centric tasks such as activity bouts or arrhythmic episodes, but remains sensitive to duty cycling and missing data~\cite{xu2025lsm}.
Hour-level granularity captures meso-scale patterns aligned with state-level phenomena such as sleep quality or fatigue~\cite{moreau2023overview}.
Day or week-level granularity stabilizes long-term trends while increasing robustness to sporadic missing data~\cite{phillips2017irregular}. 
However, these gains come at the cost of attenuating short-lived events and obscuring fine-grained variations.

\subsection{Discussion}

\tikzstyle{my-box}= [
    rectangle,
    draw=gray,
    rounded corners,
    text opacity=1,
    minimum height=1.5em,
    minimum width=5em,
    inner sep=2pt,
    align=center,
    fill opacity=.5,
]
\tikzstyle{leaf}=[
my-box, 
minimum height=1.5em,
fill=yellow!27, 
text=black,
align=left,
font=\scriptsize,
inner xsep=5pt,
inner ysep=4pt,
align=left,
text width=45em,
]
\tikzstyle{leaf2}=[
my-box, 
minimum height=1.5em,
fill=purple!22, 
text=black,
align=left,
font=\scriptsize,
inner xsep=5pt,
inner ysep=4pt,
]
\tikzstyle{leaf3}=[
my-box, 
minimum height=1.5em,
fill=blue!57, 
text=black,
align=left,
font=\scriptsize,
inner xsep=5pt,
inner ysep=4pt,
]
\tikzstyle{leaf4}=[
my-box, 
minimum height=1.5em,
fill=green!17, 
text=black,
align=left,
font=\scriptsize,
inner xsep=5pt,
inner ysep=4pt,
]

\begin{figure*}[h]
    \centering
    \begin{forest}
    forked edges,
    for tree={
      grow=east,
      reversed=true,
      anchor=base west,
      parent anchor=east,
      child anchor=west,
      base=left,
      font=\small,
      rectangle,
      draw=black,
      rounded corners,
      align=center,
      text centered,
      minimum width=4em,
      edge+={darkgray, line width=1pt},
      s sep=3pt,
      inner xsep=2pt,
      inner ysep=3pt,
      ver/.style={
        rotate=90,
        child anchor=north,
        parent anchor=south,
        anchor=center,
        align=center,
        my-box
      },
    },
    where n children=0{align=left,text centered=false}{},
    where level=1{text width=7em,font=\scriptsize}{},
    where level=2{text width=10em,font=\scriptsize}{},
    where level=3{text width=8em,font=\scriptsize}{},
    where level=4{text width=6.4em,font=\scriptsize}{},
[
    \textbf{LSMs without language capability}, ver
        [
            \textbf{Generative Strategy}
                [
                   \textbf{Signal Reconstruction}
                        [{M-REGLE~\cite{zhou2025applying},
                          EEGFormer~\cite{chen2024eegformer}, \\
                          REGLE~\cite{yun2024unsupervised},
                          LaBraM~\cite{jiang2024large}},
                          leaf4, text width=24em %27
                        ]
                ]
                [
                    \textbf{Masked Reconstruction}
                        [{LSM~\cite{narayanswamy2024scaling},
                          \cite{mathew2024foundation},
                          PAT~\cite{ruan2025ai}, \\
                          HuBERT-ECG~\cite{coppola2024hubert},
                          NORMWEAR~\cite{luo2024toward}, \\
                          LSM-2~\cite{xu2025lsm},
                          \cite{abbaspourazad2024wearable}, 
                          MMM~\cite{yi2023learning}, \\
                          CBraMod~\cite{wang2025cbramod},
                          HeartLang~\cite{jin2025reading},\\
                          SelfPAB~\cite{logacjov2024selfpab},
                          BrainWave~\cite{yuan2024brainwave}, \\
                          Brant~\cite{zhang2023brant},
                          HeartBeiT~\cite{vaid2023foundational}},
                          leaf4, text width=24em
                        ]
                ]
                [
                    \textbf{Augmentation Reconstruction}
                        [{\cite{yuan2024self}},
                          leaf4, text width=24em
                        ]
                ]
        ]
        [
            \textbf{Contrastive Strategy}
                [
                \textbf{Sample-Level Contrastive}
                    [{Pulse-PPG~\cite{saha2025pulse},
                      SiamQuality~\cite{ding2024siamquality},
                      \cite{lai2023practical}, \\
                      RELCON~\cite{xu2024relcon}},
                      leaf4, text width=24em
                    ]
                ]
                [
                \textbf{Subject-Level Contrastive}
                    [{PAPAGEI~\cite{pillai2025papagei},
                      \cite{erturk2025beyond},
                      \cite{abbaspourazad2024largescale}},
                      leaf4, text width=24em
                    ]
                ]
                [
                \textbf{Modality-Level Contrastive}
                    [{SleepFM~\cite{thapa2024sleepfm},
                      \cite{abbaspourazad2024wearable}},
                      leaf4, text width=24em
                    ]
                ] 
                [
                \textbf{Sequence-Level Contrastive}
                    [{ECGFM~\cite{zhang2025ecgfm}},
                      leaf4, text width=24em
                    ]
                ] 
        ]
        [
            \textbf{Hybrid Strategy}
                [
                \textbf{Contrastive \& Generative}
                    [{ECG-FM~\cite{mckeen2024ecg},
                      OPERA~\cite{zhang2024towards},
                      \cite{song2024foundation}},
                      leaf4, text width=24em
                    ]
                ]
        ]
]
    \end{forest}
    \vspace{-0.1in}
    \caption{An overview of LSMs without language capability.}
    \label{fig:overview}
\end{figure*}
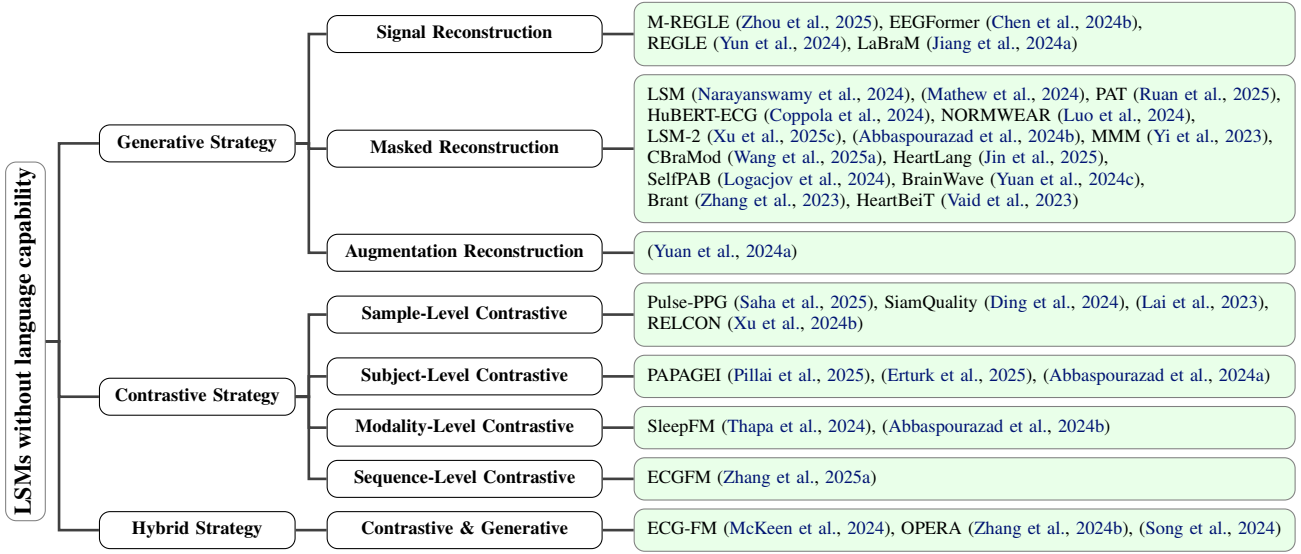

\textbf{Challenges of Wearable Data.}
Wearable data exhibits structural and operational challenges that fundamentally distinguish it from other FM domains. 
\begin{itemize}
    \item (1) Modality heterogeneity produces signals with different temporal resolutions noise characteristics, irregular sampling patterns, and long-tailed distributions~\cite{gu2025foundation, xu2024taming}, complicating joint representation learning across sensor modalities. 
    \item (2) Wearable sensing is inherently embodied: measurements depend on device placement, body morphology, and physical context, inducing distribution shifts that have no analogue in text or vision datasets~\cite{cai2025towards, singh2025feel}. 
    \item (3) Wearable data lack intuitive semantic structure and large-scale labeled corpora~\cite{xu2021limu, haresamudram2022assessing}. 
    The non-intuitive nature of raw sensor streams makes annotation costly and indirect, resulting in limited supervision and restricting the scalability and generalization of downstream tasks.
    \item (4) Modality missing is prevalent in real-world deployments, where sensors may be intermittently unavailable~\cite{narayanswamy2024scaling, zhang2022m3care}. 
    This results in incomplete multimodal observations, posing additional challenges for comprehensive representation generation and cross-modal integration.
\end{itemize}

\textbf{Modality.}
For LSMs to emerge as universal sensor FMs, their representations need to capture the distinctive temporal and statistical properties of each modality while simultaneously providing a unifying representational space across modalities.
currently, most research on LSMs has mainly concentrated on physiological or motion data.
More recent studies have begun to integrate context or position modalities~\cite{ultraposer, devrio2023smartposer, oh2024uwb, demirel2025using}, but such explorations remain fragmented and limited in scale.
To this end, a systematic investigation of context and position modalities remains unexplored yet holds significant potential, as these modalities uniquely embed relational and contextual dimensions that are not captured by physiological or motion data alone.

\textbf{Structure.}
Low-level data retain fine-grained temporal dynamics, enabling sensor encoders to learn rich representations, while high-level abstractions provide semantically grounded and more stable representations suited to interpretable or clinical-aligned applications. 
Similarly, shorter granularity windows preserve responsiveness to transient events, whereas longer horizons emphasize stable trends and reduce variance under irregular sampling. 
Therefore, flexibly leveraging multi-level granularity and abstraction, or developing adaptive mechanisms, is crucial for building unified LSMs that can balance temporal precision, semantic interpretability, and robustness across diverse tasks and deployment settings.

\section{LSMs without Language Capability}
\label{sec:modeling1}

The effectiveness of LSMs highly depends on their modeling paradigm, which affects the generalization capability of learned representations across domains and downstream tasks. 
In this section, we focus on large-scale pretraining, the dominant paradigm in current LSM research.
An summary of existing LSMs is presented in Figure~\ref{fig:overview}.

\subsection{Scaling Law}

Scaling has emerged as the most reliable route to improve performance for language and vision FMs, where model performance follows predictable power-law decay as parameters, data size, and compute increase~\cite{kaplan2020scaling, hoffmann2022training, bahri2024explaining}. 
Recently, \citet{narayanswamy2024scaling} indicate that similar regularities may extend to sensor models, reporting consistent gains in generative and discriminative tasks. 
Crucially, these models display notable sample efficiency in label-scarce settings, a property central to wearable data, where annotations are usually costly and difficult to obtain.

However, scaling in wearable sensing presents challenges absent from text and vision.
Sensor streams are inherently nonstationary, device heterogeneity, placement variability, and calibration mismatches introduce distribution shifts that may amplify rather than attenuate with scale. 
Existing demonstrations~\cite{narayanswamy2024scaling} of the sensor scaling law have been largely restricted to homogeneous hardware and the same wear locations, leaving open whether power-law dynamics persist in heterogeneous real-world deployments.
We argue that establishing the validity of scaling laws in more realistic settings is foundational for the development of LSMs.
If such regularities hold despite the embodied heterogeneity, they would provide a strong empirical justification for massively scaling LSMs as a universal substrate for wearable intelligence.

\subsection{Learning Paradigms}

Pretraining strategies shape the generalization capability of LSMs.
Existing approaches can be divided into three categories: generative, contrastive, and hybrid strategies.

\begin{figure}[t]
	\centering
	\includegraphics[width=3.3in]{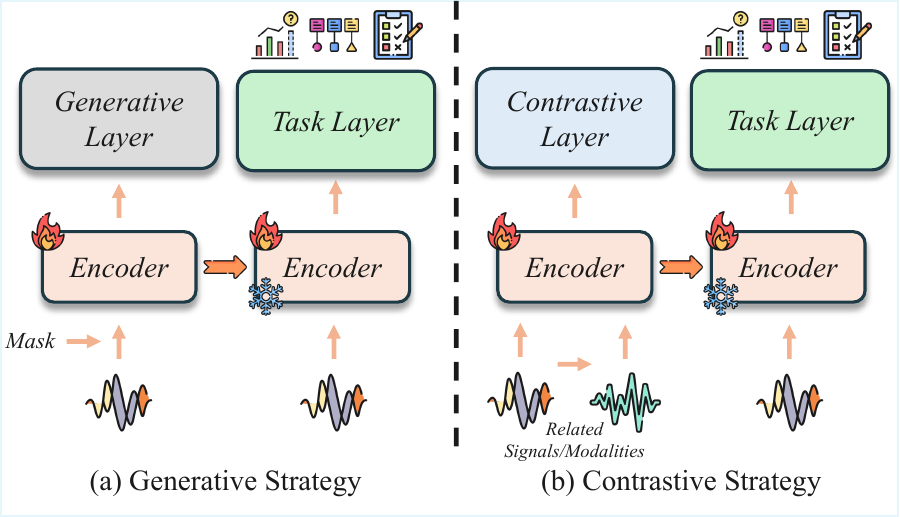}
    \vspace{-7mm}
	\caption{Illustrations of LSMs with large-scale pre-training.}
    \vspace{-5mm}
	\label{fig:modeling1}
\end{figure}

\textbf{Generative Strategy.}
Generative methods exploit the intrinsic structure of the wearable streams by reconstructing or predicting raw signals, as shown in Figure \ref{fig:modeling1} (a). 
Some methods encode and reconstruct sensor data using continuous~\cite{zhou2025applying, yun2024unsupervised} or discrete tokens~\cite{chen2024eegformer, jiang2024large}, producing compact latent sensor spaces. 
Inspired by masked language modeling~\cite{devlin2019bert}, recent work has adapted masked reconstruction to sensor data~\cite{narayanswamy2024scaling, mathew2024foundation, ruan2025ai, coppola2024hubert, luo2024toward, abbaspourazad2024wearable, wang2025cbramod, jin2025reading, yuan2024brainwave, logacjov2024selfpab, yi2023learning, zhang2023brant, vaid2023foundational}, encouraging models to capture contextual and cross-channel dependencies. 
\citet{xu2025lsm} further integrate natural missingness with artificial masking to enhance robustness under irregular sampling, while \citet{yuan2024self} propose augmentation-aware objectives that promote invariance to common signal transformations.

\textbf{Contrastive Strategy.}
Contrastive learning has emerged as a central paradigm for extracting invariant structure from heterogeneous wearable streams.
The core idea is to maximize the agreement between semantically aligned representations while discouraging spurious correlations across unrelated samples, as shown in Figure \ref{fig:modeling1} (b). 
Mainstream contrastive formulations in wearable modeling can be categorized into four granularity levels: sample-level~\cite{saha2025pulse, ding2024siamquality, xu2024relcon, lai2023practical}, subject-level~\cite{pillai2025papagei, erturk2025beyond, abbaspourazad2024largescale}, modality-level~\cite{thapa2024sleepfm, zhang2025sensorlm, tian2024foundation}, and sequence-level~\cite{zhang2025ecgfm} alignment.

\textbf{Hybrid Strategy.}
Hybrid strategies integrate generative and contrastive objectives to harness their complementary advantages. 
Joint optimization methods have been shown to generate representations that capture both local and global sensor patterns~\cite{song2024foundation, mckeen2024ecg, zhang2024towards}. 

\subsection{Discussion}

Current modeling strategies in wearable sensing underscore the difficulties of directly transferring paradigms from language or vision. 
Generative strategies impose structured latent spaces through reconstruction, yet their emphasis on data fidelity can entangle sensor noise with underlying details~\cite{oord2018representation}, limiting robustness across hardware and sampling regimes.
Contrastive strategies promote feature invariance across data transforms, subjects, or modalities, but their effectiveness heavily depends on pair construction~\cite{tian2020makes} and is susceptible to misalignment~\cite{chuang2020debiased}, which constrains their general applicability in heterogeneous real-world settings.
Hybrid strategies attempt to reconcile these objectives, yet the fundamental problems mentioned above remain unresolved~\cite{mckeen2024ecg}.

We posit that representation learning for wearable sensing should be centered on the intrinsic physical properties of sensors and the unique scaling laws governing sensor data, while explicitly disentangling embodied, distribution-specific factors, such as device characteristics and placement variability.
By embedding physical invariances, scaling-compliant design, and embodied constraints as core inductive biases, LSMs can move beyond borrowed methodological templates and evolve into a principled FM paradigm tailored to the complexities of embodied human sensing.

\section{LSMs with Language Capability}
\label{sec:modeling2}

\tikzstyle{my-box}= [
    rectangle,
    draw=gray,
    rounded corners,
    text opacity=1,
    minimum height=1.5em,
    minimum width=5em,
    inner sep=2pt,
    align=center,
    fill opacity=.5,
]
\tikzstyle{leaf}=[
    my-box,
    minimum height=1.5em,
    fill=yellow!27,
    text=black,
    align=left,
    font=\scriptsize,
    inner xsep=5pt,
    inner ysep=4pt,
    text width=45em,
]
\tikzstyle{leaf2}=[
    my-box,
    minimum height=1.5em,
    fill=purple!22,
    text=black,
    align=left,
    font=\scriptsize,
    inner xsep=5pt,
    inner ysep=4pt,
]
\tikzstyle{leaf3}=[
    my-box,
    minimum height=1.5em,
    fill=blue!57,
    text=black,
    align=left,
    font=\scriptsize,
    inner xsep=5pt,
    inner ysep=4pt,
]
\tikzstyle{leaf4}=[
    my-box,
    minimum height=1.5em,
    fill=green!17,
    text=black,
    align=left,
    font=\scriptsize,
    inner xsep=5pt,
    inner ysep=4pt,
]

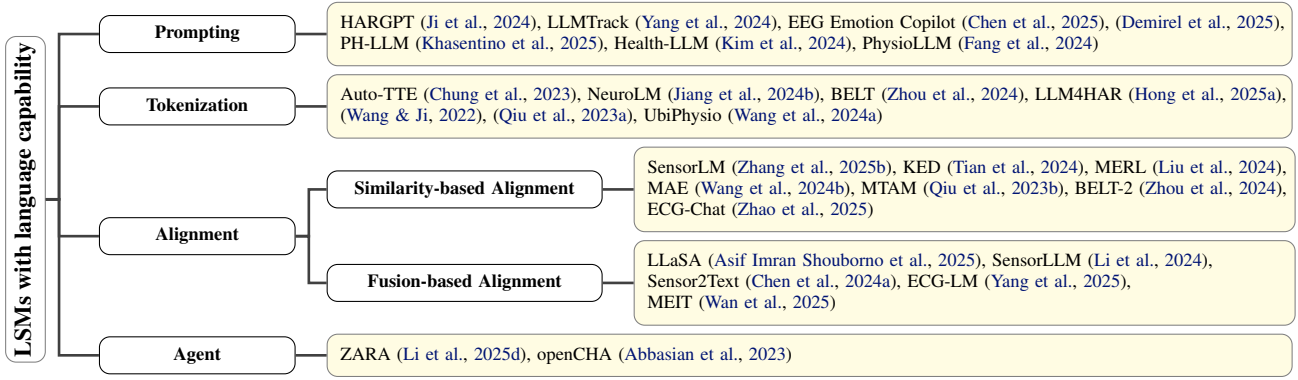
\begin{figure*}[h]
\centering
\begin{forest}
forked edges,
for tree={
    grow=east,
    reversed=true,
    anchor=base west,
    parent anchor=east,
    child anchor=west,
    base=left,
    font=\small,
    rectangle,
    draw=black,
    rounded corners,
    align=center,
    text centered,
    minimum width=4em,
    edge+={darkgray, line width=1pt},
    s sep=3pt,
    inner xsep=2pt,
    inner ysep=3pt,
    ver/.style={
        rotate=90,
        child anchor=north,
        parent anchor=south,
        anchor=center,
        align=center,
        my-box
    },
},
where n children=0{align=left, text centered=false}{},
where level=1{text width=7em, font=\scriptsize}{},
where level=2{text width=10em, font=\scriptsize}{},
where level=3{text width=8em, font=\scriptsize}{},
where level=4{text width=6.4em, font=\scriptsize}{},
[
    \textbf{LSMs with language capability}, ver
    [
        \textbf{Prompting}
        [{HARGPT~\cite{ji2024hargpt},
        LLMTrack~\cite{yang2024you},
        EEG Emotion Copilot~\cite{chen2025eeg},
        \cite{demirel2025using},\\
        PH-LLM~\cite{khasentino2025personal},
        Health-LLM~\cite{kim2024health},
        PhysioLLM~\cite{fang2024physiollm}}, leaf, text width=35.5em]
    ]
    [
        \textbf{Tokenization}
        [{Auto-TTE~\cite{chung2023text},
        NeuroLM~\cite{jiang2024neurolm},
        BELT~\cite{zhou2024belt},
        LLM4HAR~\cite{hong2025llm4har},\\
        \cite{wang2022open},
        \cite{qiu2023transfer},
        UbiPhysio~\cite{wang2024ubiphysio}}, leaf, text width=35.5em]
    ]
    [
        \textbf{Alignment}
        [
            \textbf{Similarity-based Alignment}
            [{SensorLM~\cite{zhang2025sensorlm},
            KED~\cite{tian2024foundation},
            MERL~\cite{liu2024zeroshot},\\
            MAE~\cite{wang2024enhancing},
            MTAM~\cite{qiu2023can},
            BELT-2~\cite{zhou2024belt},\\
            ECG-Chat~\cite{zhao2025ecg}}, leaf, text width=24em]
        ]
        [
            \textbf{Fusion-based Alignment}
            [{LLaSA~\cite{asif2025llasa},
            SensorLLM~\cite{li2024sensorllm},\\
            Sensor2Text~\cite{chen2024sensor2text},
            ECG-LM~\cite{yang2025ecg},\\
            MEIT~\cite{wan2025meit}}, leaf, text width=24em]
        ]
    ]
    [
        \textbf{Agent}
        [{ZARA~\cite{li2025zara},
        openCHA~\cite{abbasian2023conversational}}, leaf, text width=35.5em]
    ]
]
\end{forest}
\vspace{-0.1in}
\caption{An overview of LSMs with language capability.}
\label{fig:overview2}
\end{figure*}

LLMs have demonstrated emergent abilities beyond text to understand the physical world~\cite{xu2024penetrative, lecun2022path, guo2025language}. 
This raises a foundational question for wearable AI: can LLMs meaningfully interpret multimodal sensor data and function as a semantic interface for continuous human-centric data?
The primary obstacle lies in bridging the modality gap between discrete languages and continuous, noisy, and temporally structured sensor data.
Unlike text, wearable data lack inherent token boundaries and explicit semantic units, which makes direct integration with LLMs nontrivial.
Effective incorporation therefore requires principled mechanisms that render sensor signals compatible with language understanding while preserving physical meaning.
From this perspective, we identify four directions for expanding LLMs to LSMs: (i) prompting, (ii) tokenization, (iii) alignment, and (iv) agent.

\subsection{Prompting}

\begin{figure}[t]
	\centering
	\includegraphics[width=3.3in]{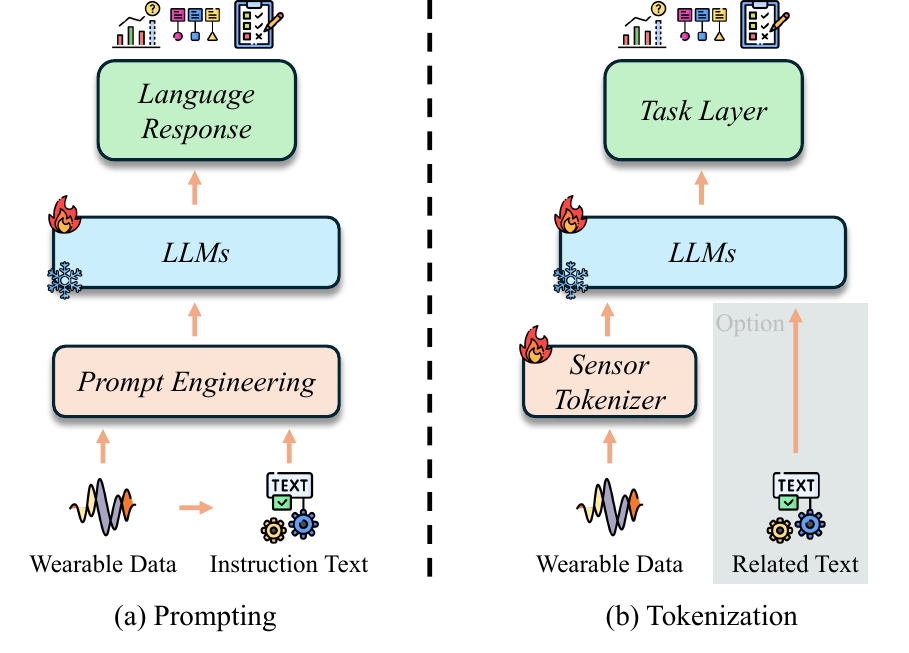}
    \vspace{-7mm}
	\caption{Illustrations of prompting and tokenization.}
	\label{fig:llm1}
\end{figure}

Prompting methods apply LLMs directly to raw sensor streams by serializing wearable time-series into textual form, treating them as pseudo-tokens, as illustrated in Figure~\ref{fig:llm1}(a). 
This strategy aims to leverage the pretrained knowledge and reasoning capacity of LLMs through carefully designed prompts that guide interpretation of sensor data. 
HARGPT~\cite{ji2024hargpt} and LLMTrack~\cite{yang2024you} explore zero-shot task adaptation using LLMs for specific downstream tasks, while EEG Emotion Copilot~\cite{chen2025eeg} employs EEG-oriented LLMs to recognize emotion states. 
\citet{demirel2025using} employ LLMs for the late fusion of motion and context sensor data for HAR.
PH-LLM~\cite{khasentino2025personal}, Health-LLM~\cite{kim2024health}, and PhysioLLM~\cite{fang2024physiollm} leverage multimodal senosr data for personalized health modeling. 

\subsection{Tokenization}

Tokenization methods transform continuous wearable data into token representations that can be directly processed by LLMs, as shown in Figure \ref{fig:llm1} (b). 
The tokenizer plays a central role, as it determines how the temporal structure and semantics are preserved when mapping sensor streams to a language-compatible token space.

Auto-TTE~\cite{chung2023text}, NeuroLM~\cite{jiang2024neurolm} and BELT~\cite{zhou2024belt} tokenize physiological data into discrete formats to perform generation and classification tasks with text instructions. 
LLM4HAR~\cite{hong2025llm4har} employs a sensor adaptation module that converts IMU data to text formats, while \citet{wang2022open} design an EEG-to-Text decoder to generate text tokens.
\citet{qiu2023transfer} introduce an optimal transport distance to achieve a fine-grained match between the ECG data and the textual annotations.
UbiPhysio~\cite{wang2024ubiphysio} introduces an activity descriptor that converts motion sequences into activity tokens, which are then fused with textual descriptions to generate professional feedback from LLMs. 

\subsection{Alignment}

\begin{figure}[t]
	\centering
	\includegraphics[width=3.3in]{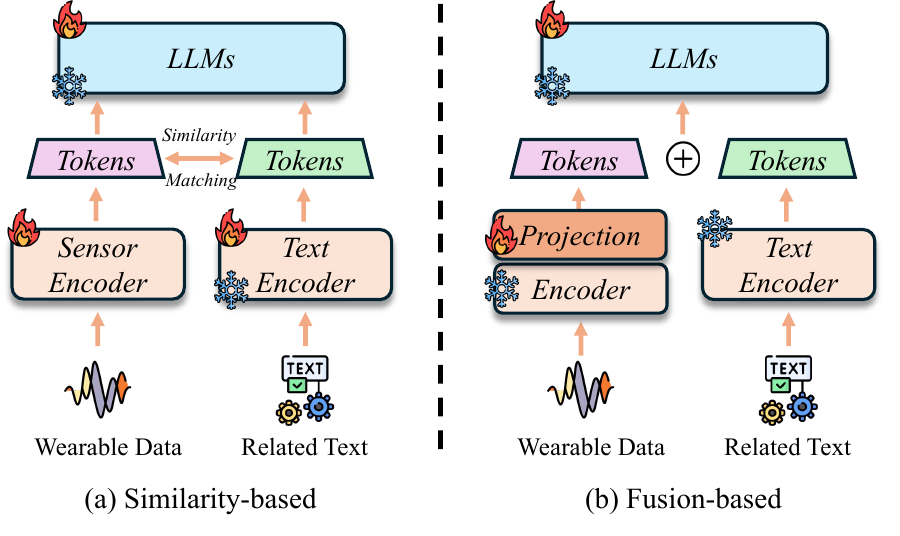}
    \vspace{-7mm}
	\caption{Illustrations of Aligning.}
	\label{fig:llm2}
    \vspace{-5mm}
\end{figure}

Alignment methods employ dedicated encoders for sensor and language modalities, projecting their representations into a shared semantic space where sensor semantics can be directly associated with textual concepts.
Existing work can be categorized into two paradigms: (i) similarity-based alignment and (ii) fusion-based alignment.

\textbf{Similarity-based Alignment.} 
Similarity-based alignment constructs paired sensor–language data and optimizes representational proximity within a shared embedding space, as shown in Figure \ref{fig:llm2} (a). 
SensorLM~\cite{zhang2025sensorlm}, KED~\cite{tian2024foundation}, and ECG-Chat~\cite{zhao2025ecg} augment textual descriptions and employ contrastive objectives to align sensor embeddings with semantic representations.
MERL~\cite{liu2024zeroshot} extends this paradigm to zero-shot ECG classification.
BELT-2~\cite{zhou2024belt} employ byte-pair encoding to facilitate the alignment between text and EEG data, while CET-MAE~\cite{wang2024enhancing} aligns text and EEG leveraging the similarity between the masked text and sensor data.
MTAM~\cite{qiu2023can} further explores EEG–text alignment using Canonical Correlation Analysis and Wasserstein Distance.

\textbf{Fusion-based Alignment.}
Fusion-based alignment inserts projection modules after the sensor encoder to align sensor embeddings and text embeddings in the input space of LLMs, as shown in Figure \ref{fig:llm2} (b). 
LLaSA~\cite{asif2025llasa} and SensorLLM~\cite{li2024sensorllm} adopt this paradigm for downstream HAR.
Sensor2Text~\cite{chen2024sensor2text} further incorporates the vision modality to expand question-answering capabilities.
ECG-LM~\cite{yang2025ecg} applies this framework to cardiovascular disease detection and ECG-related question answering, while MEIT~\cite{wan2025meit} leverages similar fusion mechanisms for automated ECG report generation.

\subsection{Agent}

Beyond serving as a central processor for sensor data, LLMs can also function as agents within wearable sensing pipelines. 
In this paradigm, LLMs coordinate modular components to plan, retrieve, and reason over multimodal information. 
OpenCHA~\cite{abbasian2023conversational} integrates an orchestrator to gather external knowledge and generate personalized responses for healthcare queries. 
ZARA~\cite{li2025zara} adopts a hierarchical multi-agent reasoning framework to perform zero-shot activity recognition.

\subsection{Discussion}

LSMs with language capability have achieved notable progress by leveraging the reasoning and semantic capabilities of LLMs. 
However, current methods focus mainly on single sensor modalities, such as IMU, ECG, or EEG, and are constrained by the scarcity of large-scale paired sensor–text corpora~\cite{asif2025llasa}. 
Moreover, the substantial computational and memory overhead of LLM integration limits deployment in real-world settings, while text-mediated reasoning pipelines may exhibit instability when applied to noisy sensor streams~\cite{park2025revisiting}. 
Therefore, future language-enabled LSMs should prioritize scalable multimodal tokenization and alignment, cost-effective adaptation, the development of robust and reliable sensor agents, and the mitigation of hallucinations.

\section{Application Areas}
\label{sec:application}

LSMs create pathways from sensor data streams to mission-critical applications. 
We highlight five potential areas where the adoption of LSMs can generate a distinctive impact.

\subsection{Personal Assistant}

Traditional personalized assistants rely on self-reported inputs and periodic logs, typically operating on coarse-grained lifestyle information~\cite{knight2021mobile}. 
In contrast, LSMs offer a unique advantage by continuously integrating user historical data with real-time multimodal sensor data. 
This enables proactive assistance, such as automatic body tracking~\cite{truslow2024understanding}, inference of health and recovery states~\cite{daniore2024wearable, fang2024physiollm}, and personalized recommendations~\cite{khasentino2025personal, abbasian2023conversational}. 
By leveraging fine-grained sensor data, LSMs have the potential to provide more timely, adaptive, and context-aware support, potentially surpassing human-level expertise in certain domains~\cite{khasentino2025personal}.

\subsection{Healthcare and Monitoring}

Wearable devices continuously collect wearable data that are valuable for disease detection and diagnosis. 
Existing studies have shown that specific sensor modalities can support the detection of particular health conditions~\cite{ballinger2018deepheart, pedroso2025leveraging, yang2022artificial, brasier2024applied}. 
Beyond detection, wearable data enable longitudinal monitoring of daily body states, facilitating early warning, risk detection, and accident prevention~\cite{li2025advancing, abedi2024artificial, adans2020enabling, kim2025simplified}. 
By integrating multimodal data within a unified representational framework, LSMs have the potential to consolidate these fragmented efforts and enable comprehensive, real-time health monitoring at scale.

\subsection{Fitness and Lifestyle}

Wearable devices have become integral to fitness and lifestyle monitoring, supporting downstream tasks such as HAR~\cite{yuan2024self, chan2024capture}, sleep analysis~\cite{birrer2024evaluating, zhai2024challenges, lim2024accurately, zheng2024sleep, khasentino2025personal}, dietary tracking~\cite{romero2023ai4fooddb, bell2020automatic}, and emerging behavior patterns that were previously underexplored~\cite{stamatakis2022association}. 
LSMs can enable comprehensive daily monitoring systems with unified multimodal representations that move beyond generic lifestyle recommendations toward adaptive, professional feedback, supporting sustained behavioral and lifestyle improvement.

\subsection{Embodied Intelligence}

By capturing comprehensive wearable information of human intent and context, LSMs can provide rich and fine-grained cues for embodied intelligence. 
LSMs enable tight coupling between users and their surroundings, thus improving intent-aware and adaptive interaction intelligence~\cite{moin2021wearable, hiremath2022bootstrapping, kaifosh2025generic, gromov2019proximity, grandi2025virtual}. 
Furthermore, LSMs can support assistive embodied intelligence~\cite{xia2024shaping, gao2025wearable} and physically grounded embodied intelligence~\cite{mengaldo2022concise, bartolozzi2022embodied, li2025ai}.

\subsection{Emotion Support}

LSMs have the potential to enable continuous monitoring of emotional states under naturalistic, free-living conditions. 
By integrating multimodal wearable data, they can support robust affect inference beyond controlled laboratory settings~\cite{rahmani2024emowear, kang2023k, saganowski2022emognition}. 
Furthermore, LSMs can extend from passive recognition to adaptive regulation, providing timely detection or feedback for stress~\cite{pei2026quantitative}, anxiety~\cite{elgendi2026wearable}, and depression~\cite{abd2023systematic}, thus supporting proactive mental well-being~\cite{lee2024examining, chen2025eeg}.

\subsection{Transportation and Mobility}

LSMs can model long-horizon behavioral patterns, enabling a deep understanding of how individuals move through environments and how mobility interacts with body states. 
Core applications include transportation mode recognition~\cite{wang2025summary, siraj2020upic, hwang2024transportation}, road surface condition detection~\cite{hong2024smallmap, chen2016city}, location-based services~\cite{yoshimura2022acceleration, xu2025experience, hong2025experience, hong2024nationwide, wang2024courirl}, and broader analyzes of mobility-related social problems~\cite{bosch2025travel, gao2024tracing, simini2021deep}. 
By integrating multimodal sensor representations on extended timescales, LSMs can support more context-aware and human-centric transportation and mobility intelligence.

\section{Further Considerations}

In this section, we consider potential questions and examine practical challenges of LSMs that must be addressed for their responsible and scalable adoption.

\subsection{Development of LSMs}

\textbf{Data Size and Diversity.}
Although large-scale wearable datasets have been developed (see Table~\ref{tab:dataset}), they are typically restricted to a single modality or sensor type. 
In contrast, wearable AI faces a significant shortage of large-scale multimodal sensor datasets. 
Considering the challenges of wearable data discussed in Section~\ref{sec:data}, this limitation poses a serious threat to the generalization ability of LSMs. 
Therefore, developing large-scale wearable datasets that encompass diverse sensing modalities is an urgent need, as it enable a more comprehensive representation of human states and environmental contexts.
Another promising direction is the development of data generation techniques, such as cross-modal generation~\cite{bhalla2021imu2doppler} and cross-distribution generation~\cite{lu2025intrinsiccontrolnet}, to alleviate data scarcity and improve model robustness.

\textbf{Evaluation.}
Careful evaluation is essential for assessing the performance of LSMs.
Some existing approaches~\cite{thapa2024sleepfm, coppola2024hubert} evaluate LSMs using metrics from specific downstream tasks. 
However, such evaluations provide only a limited view of model capability, as they fail to capture performance across the broader range of potential applications. 
Other studies~\cite{narayanswamy2024scaling} adopt pretraining-based metrics, which partially address this limitation by offering a more general assessment of representation quality. 
In addition, fundamental questions remain regarding the capabilities, underlying mechanisms, and interpretability of FMs~\cite{lin2025survey,singh2024rethinking}. 
These highlight the need to establish fair and standardized development and evaluation frameworks that can systematically assess the performance, robustness, and internal processes of LSMs.

\subsection{Deployment of LSMs}

\textbf{Computational Cost.}
Due to the mobile and embedded nature of wearable systems, LSMs may be expected to operate in resource-constrained environments. 
However, most existing LSMs rely on large-scale architectures and substantial computational resources during training and inference. 
This mismatch raises significant deployment challenges for real-world wearable scenarios. 
One promising direction is to develop deployment-oriented FM architectures~\cite{mehta2022mobilevit, yuan2024mobile}.
Another direction is to design efficiency optimization strategies~\cite{li2020train, papa2024survey, xu2025resource}.
For example, \citet{li2020train} show that highly compressed large models can achieve better performance than inherently small models.

\textbf{Privacy.}
Due to the sensitive nature of wearable data, which capture personal physiological and behavioral information, training LSMs on such data introduces security and privacy risks~\cite{shandhi2024assessment}. 
Therefore, implementing safeguards against data leakage and misuse is essential. 
This requirement is also closely aligned with the earlier discussion on disentangling intrinsic behavioral representations from embodied attributes.
Federated learning presents a promising direction for preserving wearable data privacy by enabling decentralized training without requiring raw data sharing~\cite{hao2019efficient, kuo2025distributed}.

\section{Conclusion}
This position paper focuses on wearable AI. 
Through a systematic examination of the data substrate, modeling paradigms, and application areas of LSMs, we advance a central position: building LSMs over multimodal wearable sensing offers a principled path toward a unified framework for human-centric AI, whether equipped with language capability or not. 
We further analyze the limitations of current LSM research and outline key opportunities for future studies. 
By articulating this position, we aim to draw broader attention to the LSM paradigm and try to provide a conceptual blueprint to guide the next stage of wearable AI.

\section*{Impact Statement}

This position paper explores the development of FMs for multimodal wearable sensing and proposes guiding principles for their design. 
By conceptualizing LSMs as unified sensor backbones, we envision a broad impact across diverse human-centric daily and industry application domains. 
Ethically, we emphasize the importance of the privacy-compliant use of wearable data and transparency in the development and deployment of models.
We hope that this work stimulates further research toward the next generation of sensor models that are more generalizable, scalable, and aligned with ethical and societal considerations.

\bibliography{reference}
\bibliographystyle{icml2025}

\end{document}